\documentclass[useAMS]{mn2e}
\usepackage{times, graphicx}

\title[P Cygni iron line profile in NLS1s] {The iron K feature in Narrow Line Seyfert 1's: evidence for a P Cygni profile?}

\author[C. Done, M. A. Sobolewska, M. Gierli{\'n}ski \& N. J. Schurch]{Chris Done$^1$, Ma{\l}gorzata A. Sobolewska$^1$, Marek Gierli{\'n}ski$^{1,2}$ and Nicholas J. Schurch$^1$\\
$^1$Department of Physics, University of Durham, South Road, Durham DH1 3LE, UK\\
$^2$ Obserwatorium Astronomiczne Uniwersytetu Jagiello{\'n}skiego, 30-244 Krak{\'o}w, Orla 171, Poland}

\date{Submitted to MNRAS}
\pagerange{\pageref{firstpage}--\pageref{lastpage}}

\def\xmm{{\it XMM-Newton\/}}

\def\H0{{\rm ~km~s^{-1}~Mpc^{-1}}}

\def\etal{et al.~\/}

\def\la{\mathrel{\hbox{\rlap{\hbox{\lower4pt\hbox{$\sim$}}}{\raise2pt\hbox{$<$}}}}}
\def\ga{\mathrel{\hbox{\rlap{\hbox{\lower4pt\hbox{$\sim$}}}{\raise2pt\hbox{$>$}}}}}
\def\ls{\mathrel{\hbox{\rlap{\hbox{\lower4pt\hbox{$\sim$}}}\hbox{$<$}}}}
\def\gs{\mathrel{\hbox{\rlap{\hbox{\lower4pt\hbox{$\sim$}}}\hbox{$>$}}}}
\def\d25{D$_{\rm 25}$}

\def\.25{0.25 keV\thinspace}

\begin{document}

\maketitle

\label{firstpage}

\begin{abstract}

Narrow Line Seyfert 1 galaxies are generally accreting at high
fractions of the Eddington limit. They can show complex X-ray spectra,
with a strong `soft excess' below 2 keV and a sharp drop at $\sim$7
keV. There is strong evidence linking the soft excess to either
reflection or absorption from relativistic, partially ionized material
close to the black hole. The reflection models can also simultaneously
produce the 7 keV feature from fluorescent iron K$\alpha$ line
emission from the disc. Here we show that absorption can also produce
a sharp feature at 7 keV from the P Cygni profile which results from
absorption/scattering/emission of He- and H-like iron K$\alpha$
resonance lines in the wind. We demonstrate this explicitly by fitting
the iron feature seen in \xmm ~data from 1H 0707-495 to a P Cygni
profile. The resulting column and ionization required to produce this
feature are probably larger than those needed to produce the soft
excess. Nonetheless, the absorbing material could still be a single
structure with stratified ionization such as that produced by the
ionization instability.

\end{abstract}


\section{Introduction}

Eddington luminosity flows onto black holes represent extreme
accretion rates in extreme gravity. They are characteristic of the
majority of quasars at the peak of their activity at redshift $z\sim$2
(McLure \& Dunlop 2004), and of the first black holes which grow to
power the highest redshift quasars now known at $z>6$ (Fan \etal 2003;
Volonteri \& Rees 2005). Thus understanding such flows is important
cosmologically as well as astrophysically, yet local examples of these
which can be well studied, the Narrow Line Seyfert 1 galaxies (NLS1's:
Boroson 2002), show complex features which are not easy to
interpret. Their spectra are generally dominated by strong UV/soft
X-ray emission, consistent to zeroth order with expectations of an
accretion disc at around Eddington luminosity, $L_{\rm Edd}$, onto a
10$^6$-10$^8$ M$_\odot$ black hole (e.g. Turner \& Pounds
1989). However, the shape of the emission, especially at soft X-ray
energies and above is not well matched to disc models. The soft X-ray
emission rises rather smoothly to connect onto the peak disc UV
emission (Zheng \etal 1997; Czerny \etal 2003), not at all like the
abrupt soft X-ray rise expected from the Wien tail of accretion
disc. Instead it can be modelled by a separate thermal component, but
the `temperature' of this feature is remarkably constant over a large
range in black hole mass (Czerny \etal 2003; Gierli{\'n}ski \& Done
2004, hereafter GD04, Crummy \etal 2006), making it unlikely to be
truly related to the disc.

\begin{figure}
\centering
\begin{minipage}{85 mm}
\centering
\vbox{
\includegraphics[width=8 cm, angle=0]{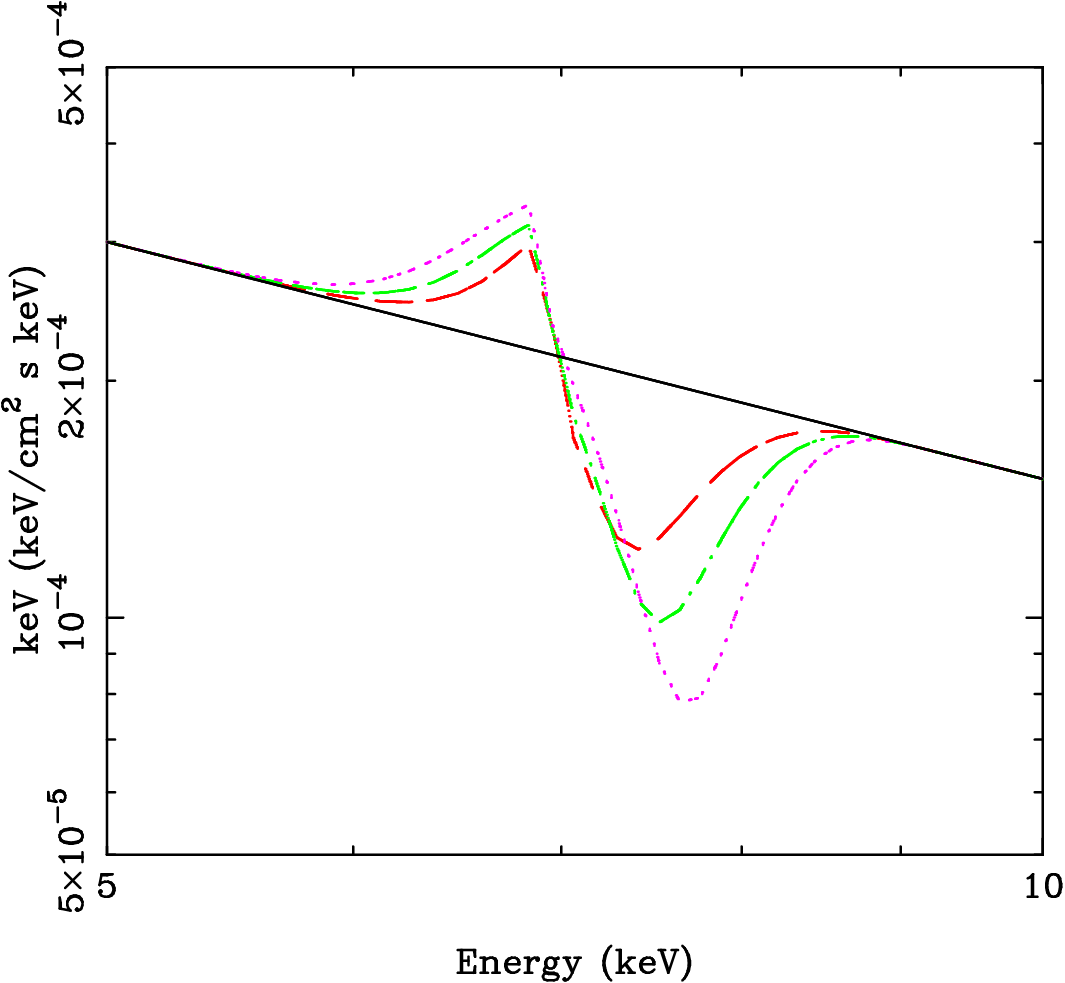}
\includegraphics[width=8 cm, angle=0]{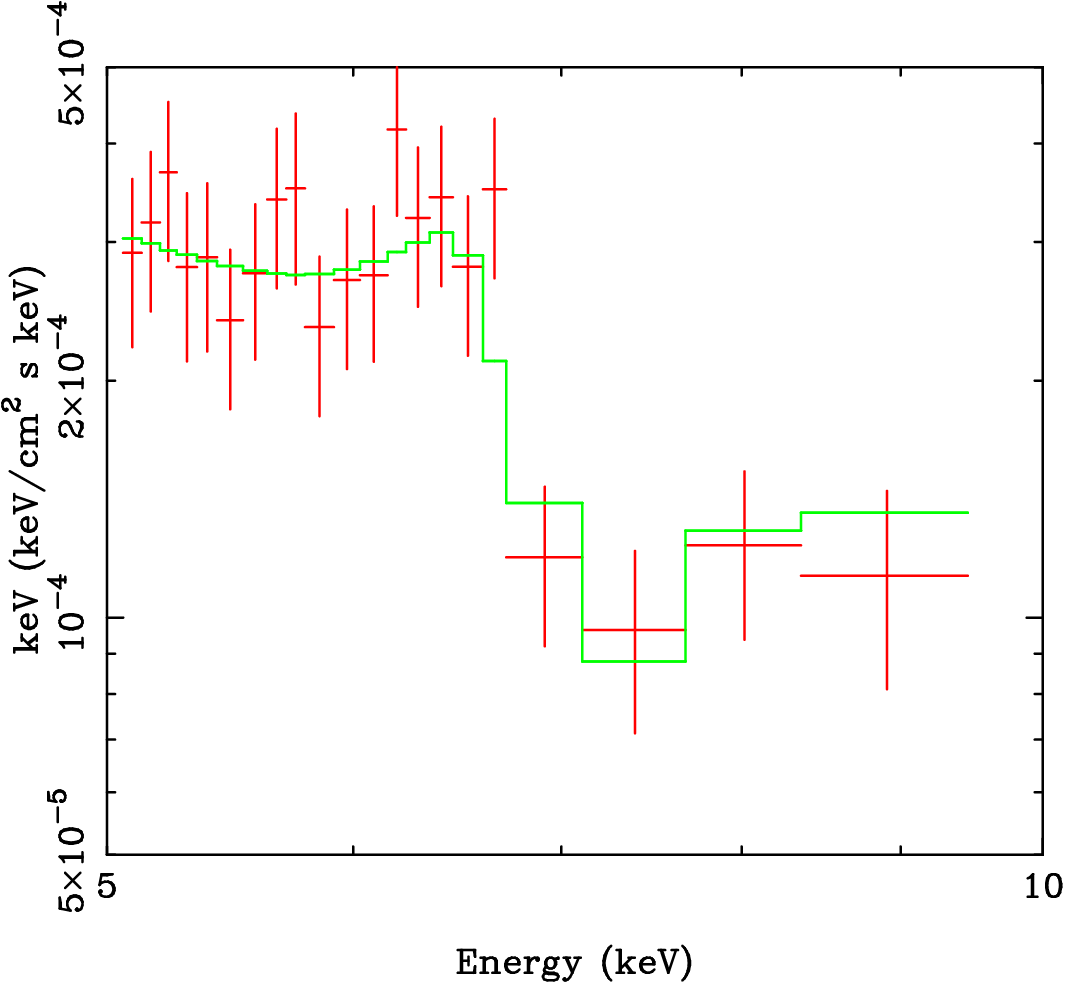}
\includegraphics[width=8 cm, angle=0]{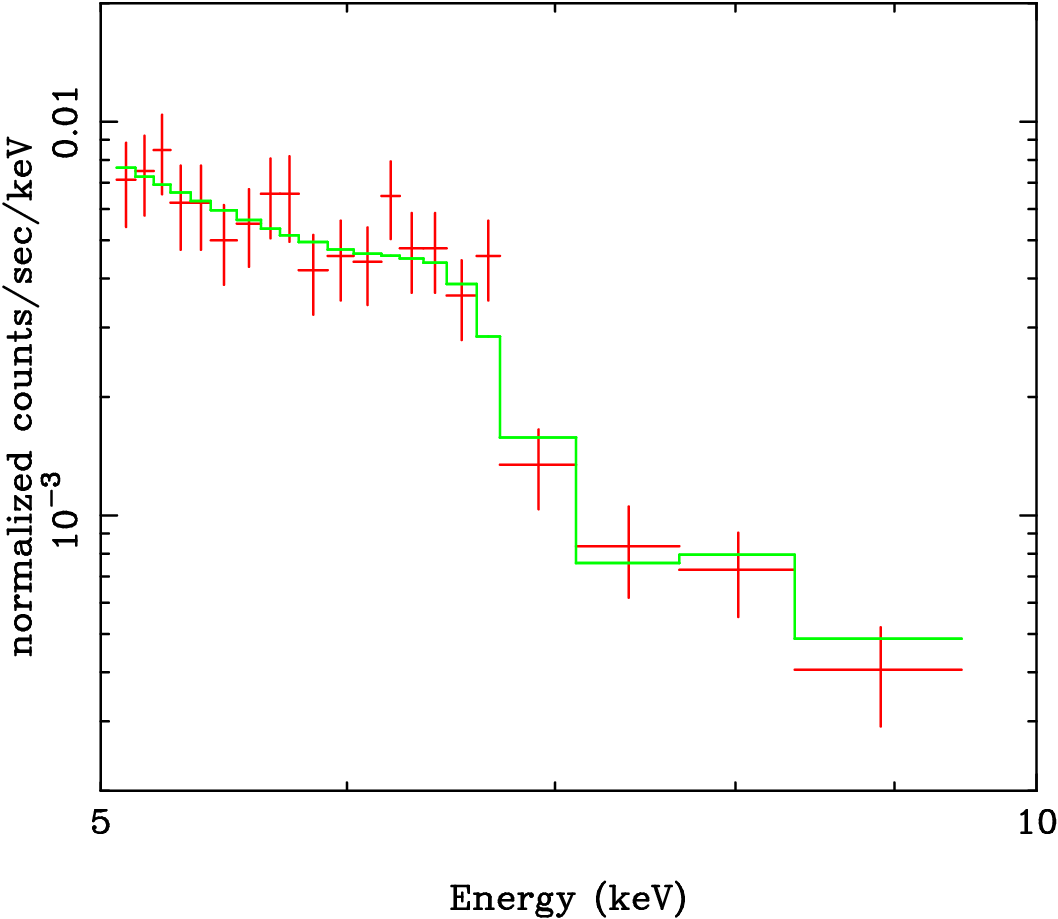}
}
\caption{{\it Upper Panel}: P Cygni profiles for the H-like resonance
iron line at 6.95 keV for a total optical depth of $\tau_{\rm
tot}$=0.5, 1, and 2 (dashed, dot-dashed and dotted curves,
respectively) using the optical depth-velocity model described in the
text. The intrinsic continuum (solid straight line) is assumed to be a
steep power law with photon index $\Gamma$=3. The sharp drop around
the rest energy of the transition is generically produced by
unsaturated absorption for material in which the distribution of
optical depth with velocity peaks at low velocities. {\it Middle
Panel}:The unfolded ($E F_E$) spectrum of 1H 0707-495 deconvolved
using the best-fitting model of a power law with $\Gamma$=3.5 and a P
Cygni profile with $\tau_{\rm tot}$=1.7 for a transition at rest
energy 6.9 keV. The data are consistent with both the finite width of 
the drop predicted by this model, and with the
recovery above 8~keV, but the limited signal--to--noise means that
neither of these features is significantly detected. {\it Lower
Panel}: The corresponding counts spectrum}
\label{fig:pcyg}
\end{minipage}
\end{figure}
 
Instead, the fixed energy is more easily explained with atomic
processes. In particular, there is an abrupt increase in opacity in
partially ionized material between $\sim$0.7-1 keV due to O{\small
VII}/O{\small VIII} and Fe transitions. This produces a large increase
in reflected or transmitted flux below 0.7 keV, and both reflection
and absorption models using partially ionized material can fit the
properties of the soft X-ray excess (Fabian \etal 2002, 2004; GD04;
Crummy \etal 2006; Chevallier \etal 2006; Schurch \& Done 2006). Both
geometries also require large velocity smearing, to hide the
characteristic {\em sharp} atomic features, where at least some of the
material is moving at high velocity $\ga$0.3c. Such speeds are
naturally produced only close to the black hole, so both these models
predict that the soft excess arises in regions of strong
gravity. However, the different models give very different pictures of
the geometry. For the reflection model, the strongest soft excesses
require that the spectrum is dominated by reflection ({\it e.g.}
Fabian \etal 2002), whereas the absorption model produces the observed
range in strength of the soft excess by simply changing the column of
material in the line of sight (Chevallier \etal 2006; Schurch \& Done
2006). Extreme spin and perhaps even direct tapping of the spin energy
of the black hole are implied by the large velocities in the
reflection model, as these are associated with a Keplerian disc
(Fabian \etal 2004, 2005; Miniutti \& Fabian 2004). By contrast, the
absorption model has the material in a wind, and its velocity shear
corresponds to the accelerating wind rather than the disc, giving no
direct constraints on the properties of the space-time.

Both reflection and absorption models give comparably good fits to the
spectra of the soft excess, and its spectral variability (Ponti \etal
2006; Gierli{\'n}ski \& Done 2006). There is a further clue to its
origin, which is that the strongest soft excesses are often seen
together with another puzzling spectral feature, namely a strong sharp
drop at $\sim$7 keV (e.g. Tanaka, Boller \& Gallo 2005). While
this energy obviously points to an association with iron, its
properties are not easy to explain (e.g Boller \etal 2002;
Gallo 2006). Atomic features of iron are expected to accompany atomic
models of the soft excess, yet the drop is surprisingly sharp for
partially ionized material in a strong gravitational field. The
reflection models can produce this feature from Doppler boosting of
the blue wing of an iron line emitted from a Keplerian disc (Fabian
\etal 1989), and a key point favouring this interpretation is that
continuum reflection from the same material can match the shape of the
soft excess ({\it e.g.} Crummy \etal 2006). However, objects with the
best signal-to-noise spectra require a more complex mix of reflection
ionizations and smearing. The soft excess shape and smooth curvature
redward of the sharp drop implies such extreme smearing that the
predicted blue wing of the line is not that sharp, so there needs to
be additional reflection from less relativistic material to match the
detailed shape of the spectrum at 7 keV ({\it e.g.} Fabian \etal
2002).

By contrast, the absorption model as currently implemented produces
only {\em broad} absorption features due to the assumption of Gaussian
(random turbulence) velocity smearing (GD04; Sobolewska \& Done 2006;
Schurch \& Done 2006). Here instead we use a structured velocity field
for the outflow, and model the resulting
emission/absorption/scattering of a strong resonance line in the
wind. The sharp feature at $\sim$7 keV is well fit by a P Cygni
profile from He- or H-like iron line from an accelerating, outflowing
wind.

\section{P Cygni profile modelling}

We use the Lamers, Cerruti-Sola \& Perinotto (1987) method of
calculating the P Cygni line profile from a spherically symmetric
wind. The resulting profile (scaled to the terminal velocity
$v_{\infty}$) is not very sensitive to the assumed velocity field
(Castor \& Lamers 1979; Lamers \etal 1987) so we pick $w \equiv
v/v_{\infty} w_{0}+(1-w_{0})(1-1/x)$, where $x=r/R_{*}$ is radial
distance $r$ in terms of the photospheric radius $R_*$, and $w_0$=0.01
is the scaled velocity at the photosphere. The profile is also not
very dependent on the distribution of material when the line is very
optically thick (saturated), giving maximum absorption ({\it i.e.}
minimum flux) at maximum outflow velocity (a classic P Cygni
profile). This is  {\em not} what is seen in the NLS1's.
The observed feature has minimum flux around 7 keV, which is
the rest energy of the most probable resonance line transitions (He-
and H-like iron K$\alpha$) in a P Cygni interpretation. Thus to fit
the observed profile requires that the line is unsaturated (not
optically thick everywhere) and has maximum absorption at more or less
the rest wavelength, with decreasing absorption with increasing
blueshift {\it i.e.} velocity along the line of sight. Thus we choose
to parameterize the optical depth in terms of velocity rather than
with radius as this gives the more direct link to the observed profile
shape (Lamers \etal 1987). The atlas of P Cygni profiles of Castor \&
Lamers (1979) gives examples of many different optical depth-velocity
distributions. We pick $\tau(w)=\tau_{0} (1-w)^4$ (so a total optical
depth $\tau_{\rm tot}$=0.2$\tau_{0}$) as an example of a profile which
can give a sharp transition between emission and absorption at around
the rest wavelength for moderate optical depths (Castor \& Lamers
1979). This is by no means a unique solution, we are merely using it
to illustrate the possibility of P Cygni models to produce the
observed feature at 7 keV. 

Fig.~\ref{fig:pcyg} ({\it Upper panel}) shows the P Cygni profiles
predicted with this model for H-like iron K$\alpha$ at rest energy
$E_0$~=6.95 keV for total optical depths of 0.5, 1 and 2 and
$v_{\infty}$=0.3c. The transition between emission and absorption is
typically rather sharp (by design), within $\Delta E$ of 0.5~keV of
the rest energy of the line. The absorption ends at
$(1+v_{\infty}/c)E_0 \sim 8$~keV, so the model predicts a recovery to
the intrinsic steep spectrum at this point.
We note that 'classic' P Cygni lines are seen from unsaturated iron
K$\alpha$ He-- and H--like ions in the Galactic Binary Cir X--1
(Schulz \& Brandt 2002), confirming that these are possible, though
the wind velocity inferred here is much larger. 

We demonstrate the model using \xmm ~data from 1H0707-495 (z=0.041)
taken in October 2000 (see Sobolewska \& Done 2006 for details of the
data extraction). This spectrum shows one of the strongest soft
excesses seen, together with a dramatic drop in flux at 7 keV (Boller
\etal 2002). We first concentrate on the iron K feature, fitting a P
Cygni profile only the 5-10 keV range so that the modelling is
insensitive to details of the spectral complexity at lower
energies. The best-fit model has $\tau_{\rm tot}\sim$1.7, a rest frame
line energy of $E_{0}$=6.76-7.07, consistent with either He- or H-like
iron, $v_{\infty}\sim$0.3c and $\Gamma$=3.5$\pm 0.8$
($\chi^{2}_{\nu}$=10.7/17). Fig.~\ref{fig:pcyg} ({\it Middle panel})
shows the data deconvolved using this best-fit model, showing
explicitly that this description of unsaturated
absorption/scattering/emission in a wind can match the sharp drop seen
in the data. However, the signal--to--noise in these data is rather
limited, which reduces their ability to test several key aspects of
the model. In particular, there are only two data points above
8~keV. These are plainly consistent with the predicted model spectrum,
including the recovery, but the uncertainties preclude a {\em
detection} of this characteristic feature. Likewise, around the rest
energy of the transition, the data are consistent with our particular
choice for the velocity structure, but they also would be even better
fit at the 1--2 sigma level by an even sharper transition. The counts
spectra corresponding to these data are shown in Fig.~\ref{fig:pcyg}
({\it Lower panel}). Better data
are clearly required in order to reveal the detailed shape of the
5--10~keV spectrum to give a more sensitive test of this model.

A nice additional property of the P Cygni model is that the energy at
which the sharp feature occurs can change quite easily, either by
changing the transition ({\it e.g.} He-like K$\alpha$ is at 6.7 keV
while H-like is 6.95 keV), or more generally by changing either the
geometry (since the relative weighting of the emission and absorption
components of the line profile depends on the solid angle subtended by
the wind), or wind velocity structure. Such changes in energy of the
sharp feature are observed in 1H 0707-495 (Gallo \etal 2004).

\begin{figure}
\centering
\begin{minipage}{85 mm}
\centering
\vbox{
\includegraphics[width=8 cm, angle=0]{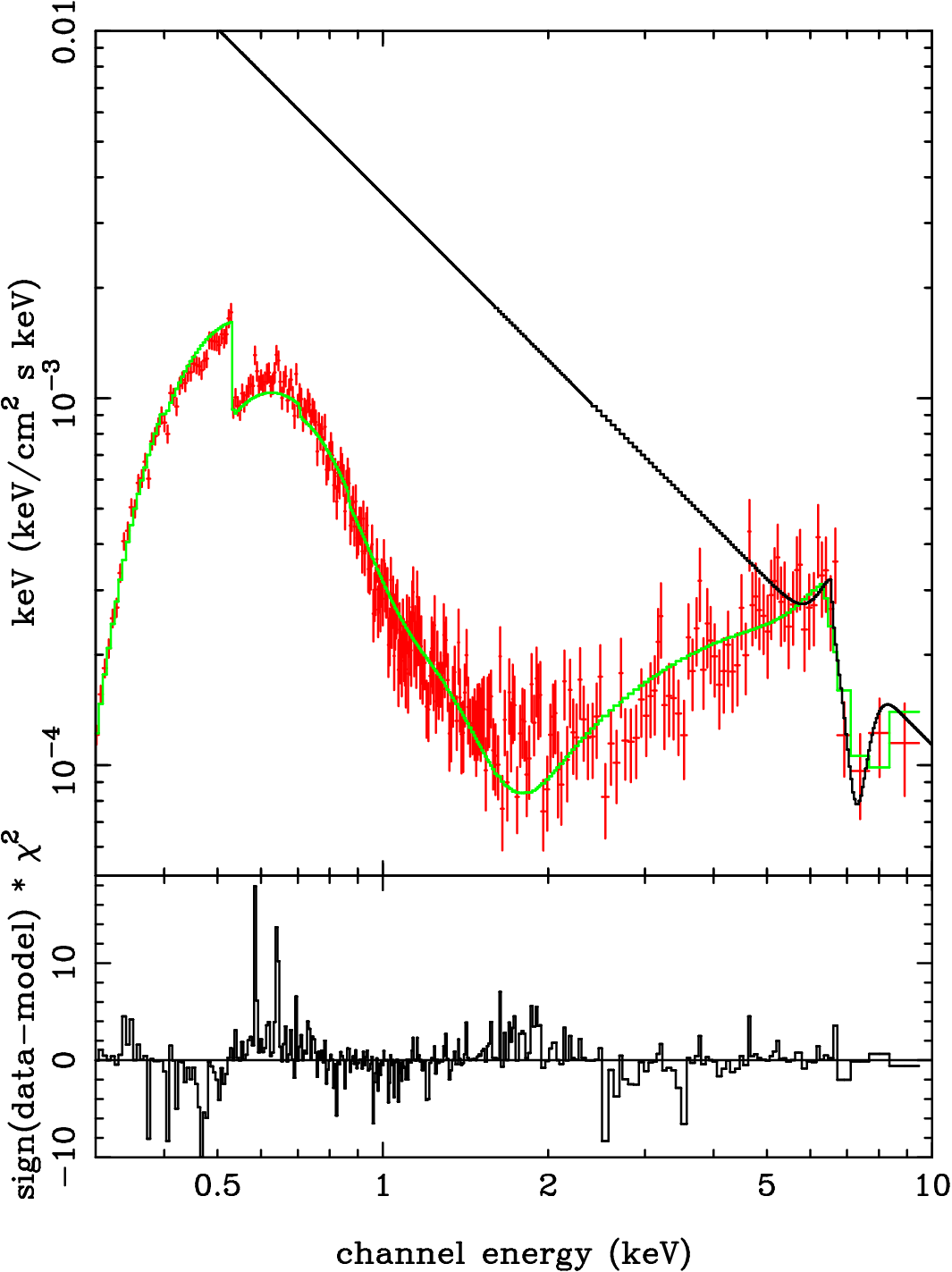}
}
\caption{The data points show the full \xmm ~spectrum of 1H 0707-495,
deconvolved with the best-fit model (solid grey curve, green in
colour) which involves a P Cygni profile for the iron features and a
smeared absorber to reproduce the continuum curvature at lower
energies (soft excess). The solid black curve shows the best-fit
intrinsic continuum inferred from fitting the 5-10 keV data with the P
Cygni profile, extrapolated over the full band. This is steep, so does
not require any intrinsic soft excess, showing that the data are
compatible with an absorption origin of this feature. The lower panel
shows residuals to the fit. Plainly there is still structure at low
energies which is not modelled here, such as emission from the wind
and/or reflection from the wind and disc and/or narrow ionized
absorption features.}
\label{fig:all}
\end{minipage}
\end{figure}

Fig.~\ref{fig:all} shows the extrapolation of the best-fitting
continuum power law inferred from these 5-10 keV fits to the lower
energy data. Plainly, the model overpredicts the softer emission,
compatible with an absorption origin of the soft X-ray excess. We
include partially ionized absorbing material with Gaussian velocity
smearing to model the lower energy curvature (see Gierli{\'n}ski \&
Done 2006 for model details), together with neutral absorption to
account for the interstellar column along the line of sight in this
galaxy and 1H 0707-495. Fig.~\ref{fig:all} shows that this composite
model (a power law modified by a P Cygni profile at iron and smeared
absorption at low energies) matches the broad shape of the overall
continuum curvature. The fit is not statistically acceptable, with
$\chi^{2}_{\nu}$=470/311, but this is unsurprising since there are
other components which are not modelled here yet which are expected to
be present at some level such as reflection/emission from the disc
and/or wind (see below) and/or narrow ionized absorption from a much
slower outflow (the classic `warm absorber'). The intrinsic,
unabsorbed flux in the 0.3--10~keV band 
inferred from this best fit is $1.4\times 10^{-11}$ ergs cm$^{-2}$
s$^{-1}$, over 6$\times$ larger than that observed (with roughly equal
amounts absorbed by the neutral gas and smeared wind). While this is a
large distortion of the X--ray flux, this is only a small fraction of
the total bolometric luminosity, which is dominated by the disc
(see the spectral deconvolution shown in Fig. 3 of GD04). 

We stress also that this is not a self-consistent model, as the iron P
Cygni profile and the soft excess absorption column are treated
differently. The iron line profile calculation includes both emission,
scattering and absorption (Sobolov approximation) in a structured
velocity field, while the soft excess column is modelled only by
absorption and a Gaussian velocity dispersion. Resonance lines in the
smeared absorption column used to make the soft excess should likewise
have P Cygni profiles, though much of the opacity at these lower
energies is from non-resonance lines and absorption edges, diluting
the P Cygni effect. A full calculation of the radiative transfer
problem in a relativistic outflow is beyond the scope of this
paper. However, we note that the derived absorption parameters for the
soft excess and iron features are probably different, suggesting that
even full radiative transfer on a single, constant ionization slab may
not match the data. This is easiest to see for the ionization
state. The `soft excess' absorber must have many transitions in the
soft X-ray region. Hence it cannot have iron predominantly as He- or
H-like, as these are so highly ionized that few lower energy
transitions should remain. The columns required are probably also
different, although the velocities are similar. The `soft excess'
column is $N_{H}$=4.6$\pm 0.2$ $\times$10$^{23}$ cm$^{-2}$ (though
this will depend on the assumed velocity structure, see Schurch \&
Done 2006). The total optical depth for the iron feature in the full
bandpass fit is 1.2$^{+5.1}_{-0.7}$, implying an ion column of
5$_{-3}^{+15}\times$10$^{19}$ cm$^{-2}$ (Lamers \etal 1987). For solar
iron abundance and a fractional He-like ion abundance of $\sim$0.3 we
estimate an equivalent Hydrogen column of
$N_{H}$=5$_{-3}^{+15}\times$10$^{24}$ cm$^{-2}$. This is optically
thick to electron scattering (see also the winds inferred by King \&
Pounds 2003).

An optically thick column does not invalidate the model, in that it
will not lead to complete thermalization of the X-ray spectrum. Such
thermalization requires that the effective optical depth (the
geometric mean of the electron scattering and true absorption optical
depth) is also larger than unity, but this is not the case here since
there is very little true absorption opacity at such high
ionization. However, it does mean that continuum scattering {\it i.e.}
reflection from the wind should also be important. This will be
smeared by the wind velocity shear, not through Keplerian motion as
for reflection from the disc, so its properties can be used to
constrain the space-time only if the wind geometry and velocity as a
function of radius are understood. However, scattering also means that
the intrinsic continuum variability will be smeared out over the size
scale of the absorber. Thus in order to produce the observed, often
dramatic variability ({\it e.g.} Leighly 1999), requires that the
absorber size scale is small assuming it subtends a large solid angle,
{\it i.e.} that the material is close to the black hole (also required
by the extreme velocity). Changes in this column could also enhance
the observed variability, perhaps producing the dramatic events seen
from these objects ({\it e.g.} Boller \etal 1997).

More importantly for this paper, the large column can also affect the
P Cygni profile region. Firstly, there should be more absorption at
higher energies, in the iron K$\beta$ and nickel K$\alpha$ lines and
in the iron edge. These should also be velocity smeared, so can mask
the characteristic recovery after the iron K$\alpha$ P Cygni line
profile (see Fig.~\ref{fig:pcyg}) and/or lead to an overestimate of
the terminal velocity. Secondly, electron scattering in
the wind can distort the P Cygni profile. Line photons are Compton up
or down scattered on collisins with free electrons, removing them from
resonance and hence from the P Cygni profile. However, this will have
the biggest effect where the line is very optically thick, so that
multiple scatterings are key to forming the line profile, whereas our
wind parameters give a resonance line optical depth around unity. Thus
we expect that electron scattering does not substantially change the
line profile, but a quantatitive assessment of all these effects
requires a proper calulation of the radiative transfer in the wind.

The inferred continuum for this model is intrinsically steep, with
$\Gamma$=3.16$\pm 0.04$. NLS1's are generally steep (Boller, Brandt \&
Fink 1996; Brandt, Mathur \& Elvis 1997), though for the obviously
curved spectrum shown here the apparent spectral index depends on the
bandpass used (Fig.~\ref{fig:all}). For these data from 1H 0707-495
the {\em apparent} index in the 2-10 keV band (with an edge to
describe the sharp drop at 7 keV) gives an extremely {\em flat}
spectrum with $\Gamma\sim$1.1 (Boller \etal 2002), while the 0.3-1 keV
band is apparently much steeper. Both reflection and absorption models
maintain that neither of these indices represent the intrinsic
illuminating spectrum, but instead result from the effects of
partially ionized material. Nonetheless, the reflection dominated fits
give slightly flatter indices, with $\Gamma$=2.5-3 (Fabian \etal 2002;
2004). Thus they do make different predictions of the high energy
flux, with the wind models giving 50--60\% less flux in the 10--30~keV
bandpass than the reflection models. This gives a potential diagnostic
test between them which may be possible with {\em Suzaku} data 
(Sobolewska \& Done 2006).

\section{Discussion}

The section above shows that the sharp drop at $\sim$7 keV in the NLS1
1H 0707-495 can be fit by an unsaturated P Cygni line profile from an
outflowing wind with total column of $\sim$5$\times$10$^{24}$
cm$^{-2}$ {\it i.e.} where the optical depth in both the resonance
line and the continuum is of order unity. We note
that this is the one of the most extreme objects in terms of the size
of the sharp drop, so other NLS1's will give lower columns. The
apparent correlation between the iron feature and the soft excess seen
at lower energies (Tanaka \etal 2005) is also naturally explained in
terms of the wind absorption, though it is likely that there is a
range of ionization parameters present in the outflowing material,
with the soft excess requiring lower column and lower ionization state
than the iron features. Even lower ionization, column and velocity
material is observed from the clear wind features present in the UV
line emission in this and other NLS1's ({\it e.g.} Leighly 2004).

Chevallier \etal (2006) present an attractive idea for a multi-ionization
column by assuming that the X-ray illuminated absorbing material is in
approximate pressure balance. This means it is subject to an
ionization instability (Krolik, McKee \& Tarter 1981). The inner face
of the cloud is highly ionized, on a stable branch where X-ray heating
is balanced by Compton cooling and bremsstrahlung. Further into the
cloud the X-ray heating is reduced, so to remain in pressure balance
requires that the density increases. Yet this increases the importance
of collisional cooling processes (bremsstrahlung and/or lines and
recombination continua) which pulls the temperature down further
(hence increasing the density). Lower temperatures mean the gas is
less highly ionized, with more ion species existing. The dramatic
increase in cooling efficiency produced by the increasing number of
atomic transitions below $\sim$0.5 keV triggers the instability where
the gas switches over a rather small range in optical depth between
being hot, highly ionized and rarified to cool, mostly neutral and
dense (Krolik \etal 1981). The ionization required in order to produce
the soft excess from atomic processes lies exactly on this instability
(Chevallier \etal 2006).

Thus this model gives a structure where there is a column containing
(at most) only He- and H-like iron, followed by a column of partially
ionized material which marks the point of the ionization
instability. Beyond this the material is mostly neutral, but the
dramatic increase in density probably causes this material to fragment
into small clumps, effectively truncating the column at this point
(Chevallier \etal 2006). Thus a single structure could produce the
apparent `soft excess' from partially ionized material across the
ionization instability, and the sharp iron feature from the highly
ionized column closer to the X-ray source. 

However, it remains to be seen whether this could work for the steep
intrinsic spectrum of $\Gamma\sim 3$ inferred here. The models of
Chevallier \etal (2006) are only calculated up to $\Gamma=2.3$, and a
steeper spectrum increases the importance of Compton cooling relative
to heating, which weakens the extent of the ionisation instability
(though does not completely suppress it e.g. 
Komossa \& Meerschweinchen 2000).

\section{Conclusions}

NLS1's are an extreme population, representing the highest $L/L_{\rm
Edd}$ AGN ({\it e.g.} Boroson 2002). They can show complex X-ray
spectra, with a strong smooth soft excess below 1 keV, and a sharp
drop around 7-8 keV. Here we show that these features can both be
produced by a wind, and that in particular the sharpness of the
feature at 7 keV matches well with a P Cygni profile produced from a
resonance line of highly ionized iron.

The alternative geometry, in which the partially ionized material is
seen in reflection, can also produce the correlated soft excess and
iron line feature ({\it e.g.} Fabian \etal 2002, 2004; Crummy \etal
2006). In this interpretation the strongest soft excess/iron features
seen require a reflection dominated geometry, and the smearing implies
an extreme space-time as the large velocity shear is interpreted as
arising from a Keplerian disc. The wind model described here also
requires some extreme parameters: the strength of the largest iron
features implies that the total column through the wind has optical
depth of order unity to electron scattering (so there should also be
reflection from the {\em wind}) and again there is a large velocity
shear. However, here the mildly relativistic terminal velocity only
implies that the wind is launched close to the black hole, rather than
giving direct constraints on the space-time.

Thus both reflection and absorption can describe the data, but both
require some extreme parameters. The question becomes which set of
extreme parameters are most physically plausible. Certainly a
reflecting disc with relativistic smearing is expected to be present
at some level. But strong winds are likewise expected from objects
with luminosity close to Eddington, especially sources with luminosity
peaking in the UV/EUV region where line driving leads to dramatic
enhancement of the radiation force (Proga, Stone \& Kallman 2000;
Proga \& Kallman 2004). Any contribution from magneto-centrifugal
forces will likewise increase the wind strength ({\it e.g.} Proga 2003
and references therein). We stress the importance of higher energy
data in distinguishing between the two geometries, and in determining
which piece of extreme physics is the most important in understanding
the spectra of these objects.

\section{Acknowledgements}

Much of this work is based on observations obtained with \xmm, an ESA
science mission with instruments and contributions directly funded by
ESA Member States and the USA (NASA). This research has made extensive
use of NASA's Astrophysics Data System Abstract Service.  CD MG \& NJS
acknowledge financial support through a PPARC Senior fellowship and
PDRF and PDRA, respectively. CD thanks ISAS for hospitality, and Aya
Kubota and Thomas Boller for encouragement. We thank the referee,
Kazushi Iwasawa, for insightful questions.

\label{lastpage}


\begin{thebibliography}{}

\bibitem[\protect\citeauthoryear{Boroson}{2002}]{2002ApJ...565...78B}
Boroson T.~A., 2002, ApJ, 565, 78.
\bibitem[\protect\citeauthoryear{Boller \etal}{2002}]{2002MNRAS.329L...1B}
Boller T., \etal, 2002, MNRAS, 329, L1.
\bibitem[\protect\citeauthoryear{Boller, Brandt, \& Fink}{1996}]{1996A&A...305...53B}
Boller T., Brandt W.~N., Fink H., 1996, A\&A, 305, 53.
\bibitem[\protect\citeauthoryear{Boller \etal}{1997}]{1997MNRAS.289..393B}
Boller T., Brandt W.~N., Fabian A.~C., Fink H.~H., 1997, MNRAS, 289, 393.
\bibitem[\protect\citeauthoryear{Brandt, Mathur, \& Elvis}{1997}]{1997MNRAS.285L..25B}
Brandt W.~N., Mathur S., Elvis M., 1997, MNRAS, 285, L25.
\bibitem[\protect\citeauthoryear{Castor \& Lamers}{1979}]{1979ApJS...39..481C}
Castor J.~I., Lamers H.~J.~G.~L.~M.,1979, ApJS, 39, 481.
\bibitem[\protect\citeauthoryear{Chevallier et al.}{2006}]{2006A&A...449..493C} 
Chevallier L., Collin S., Dumont A.-M., Czerny B., Mouchet M., Gon{\c c}alves A.~C., Goosmann R., 2006, A\&A, 449, 493.
\bibitem[\protect\citeauthoryear{Crummy \etal}{2006}]{2006MNRAS.365.1067C}
Crummy J., Fabian A.~C., Gallo L., Ross R.~R., 2006, MNRAS, 365, 1067.
\bibitem[\protect\citeauthoryear{Czerny \etal}{2003}]{2003A&A...412..317C}
Czerny B., Niko{\l}ajuk M., R\'o$\rm \dot z$a\'nska A., Dumont A.-M., Loska Z., Zycki P.~T., 2003, A\&A, 412, 317.
\bibitem[\protect\citeauthoryear{Fabian \etal}{1989}]{1989MNRAS.238..729F}
Fabian A.~C., Rees M.~J., Stella L., White N.~E., 1989, MNRAS, 238, 729.
\bibitem[\protect\citeauthoryear{Fabian \etal}{2002}]{2002MNRAS.331L..35F}
Fabian A.~C., Ballantyne D.~R., Merloni A., Vaughan S., Iwasawa K., Boller T., 2002, MNRAS, 331, L35.
\bibitem[\protect\citeauthoryear{Fabian \etal}{2004}]{2004MNRAS.353.1071F}
Fabian A.~C., Miniutti G., Gallo L., Boller T., Tanaka Y., Vaughan S., Ross R.~R., 2004, MNRAS, 353, 1071.
\bibitem[\protect\citeauthoryear{Fabian \etal}{2005}]{2005MNRAS.361..795F}
Fabian A.~C., Miniutti G., Iwasawa K., Ross R.~R., 2005, MNRAS, 361, 795.
\bibitem[\protect\citeauthoryear{Fan \etal}{2003}]{2003AJ....125.1649F}
Fan X., \etal, 2003, AJ, 125, 1649.
\bibitem[\protect\citeauthoryear{Gallo \etal}{2004}]{2004MNRAS.353.1064G}
Gallo L.~C., Tanaka Y., Boller T., Fabian A.~C., Vaughan S., Brandt W.~N., 2004, MNRAS, 353, 1064.
\bibitem[\protect\citeauthoryear{Gallo}{2006}]{2006MNRAS.368..479G} 
Gallo L.~C., 2006, MNRAS, 368, 479.
\bibitem[\protect\citeauthoryear{Gierli{\'n}ski \& Done}{2004}]{2004MNRAS.349L...7G}
Gierli{\'n}ski M., Done C., 2004, MNRAS, 349, L7.
\bibitem[\protect\citeauthoryear{Gierli{\'n}ski \& Done}{2006}]{2006MNRAS.tmpL..64G}
Gierli{\'n}ski M., Done C., 2006, MNRAS,L64.
\bibitem[\protect\citeauthoryear{King \& Pounds}{2003}]{2003MNRAS.345..657K}
King A.~R., Pounds K.~A., 2003, MNRAS,345, 657.
\bibitem[\protect\citeauthoryear{Krolik, McKee, \& Tarter}{1981}]{1981ApJ...249..422K} Krolik J.~H., McKee C.~F., Tarter C.~B., 1981, ApJ, 249, 422.
\bibitem[\protect\citeauthoryear{Komossa \& 
Meerschweinchen}{2000}]{2000A&A...354..411K} Komossa S., Meerschweinchen 
J., 2000, A\&A, 354, 411 
\bibitem[\protect\citeauthoryear{Lamers, Cerruti-Sola, \& Perinotto}{1987}]{1987ApJ...314..726L}
Lamers H.~J.~G.~L.~M., Cerruti-Sola M., Perinotto M., 1987, ApJ, 314, 726.
\bibitem[\protect\citeauthoryear{Leighly}{1999}]{1999ApJS..125..297L}
Leighly K.~M., 1999, ApJS, 125, 297.
\bibitem[\protect\citeauthoryear{Leighly}{2004}]{2004ApJ...611..125L}
Leighly K.~M., 2004, ApJ, 611, 125.
\bibitem[\protect\citeauthoryear{McLure \& Dunlop}{2004}]{2004MNRAS.352.1390M}
McLure R.~J., Dunlop J.~S., 2004, MNRAS, 352, 1390.
\bibitem[\protect\citeauthoryear{Miniutti \& Fabian}{2004}]{2004MNRAS.349.1435M}
Miniutti G., Fabian A.~C., 2004, MNRAS, 349, 1435.
\bibitem[\protect\citeauthoryear{Ponti \etal}{2006}]{2006MNRAS.368..903P}
Ponti G., Miniutti G., Cappi M., Maraschi L., Fabian A.~C., Iwasawa K., 2006, MNRAS, 368, 903.
\bibitem[\protect\citeauthoryear{Proga, Stone, \& Kallman}{2000}]{2000ApJ...543..686P} 
Proga D., Stone J.~M., Kallman T.~R., 2000, ApJ, 543, 686.
\bibitem[\protect\citeauthoryear{Proga}{2003}]{2003ApJ...585..406P} 
Proga D., 2003, ApJ, 585, 406.
\bibitem[\protect\citeauthoryear{Proga \& Kallman}{2004}]{2004ApJ...616..688P}
Proga D., Kallman T.~R., 2004, ApJ, 616, 688.
\bibitem[\protect\citeauthoryear{Schurch \& Done}{2006}]{2006MNRAS.tmp..762S}
\bibitem[\protect\citeauthoryear{Schulz \& 
Brandt}{2002}]{2002ApJ...572..971S} Schulz N.~S., Brandt W.~N., 2002, ApJ, 
572, 971 
Schurch N.~J., Done C., 2006, MNRAS, 762.
\bibitem[\protect\citeauthoryear{Sobolewska \& Done}{2006}]{}
Sobolewska M., Done C., 2006, MNRAS, submitted (astro-ph/0609223).
\bibitem[\protect\citeauthoryear{Tanaka, Boller, \& Gallo}{2005}]{2005gbha.conf..290T}
Tanaka Y., Boller T., Gallo L., 2005, gbha.conf, 290.
\bibitem[\protect\citeauthoryear{Turner \& Pounds}{1989}]{1989MNRAS.240..833T}
Turner T.~J., Pounds K.~A., 1989, MNRAS, 240, 833.
\bibitem[\protect\citeauthoryear{Volonteri \& Rees}{2005}]{2005ApJ...633..624V}
Volonteri M., Rees M.~J., 2005, ApJ, 633, 624.
\bibitem[\protect\citeauthoryear{Zheng \etal}{1997}]{1997ApJ...475..469Z}
Zheng W., Kriss G.~A., Telfer R.~C., Grimes J.~P., Davidsen A.~F., 1997, ApJ, 475, 469.

\end{thebibliography}
\end{document}